\documentclass[pre,preprint,aps,amsmath,amssymb]{revtex4}
\usepackage{graphicx}
\usepackage{epsfig}
\include{graphics}

\begin{document}
\title{Scintillation reduction by use of multiple  
Gaussian laser beams with different wavelengths}

\author{Avner Peleg and Jerome V. Moloney}
\affiliation{Arizona Center for  Mathematical Sciences, 
University of Arizona, Tucson, Arizona 85721, USA}

\begin{abstract}
We study the scintillation index of $N$ 
partially overlapping collimated lowest order Gaussian 
laser beams with different wavelengths in weak atmospheric turbulence. 
Using the Rytov approximation we calculate the initial beam 
separation that minimizes the longitudinal scintillation. 
Further reduction of the longitudinal 
scintillation is obtained by optimizing with respect to both
beam separation and spot size.
The longitudinal scintillation of the optimal 
$N$-beam configurations is inversely proportional to $N$, 
resulting in a 92$\%$ reduction 
for a 9-beam system compared with the single 
beam value. The radial scintillation values  
for the optimal $N$-beam configurations are 
significantly smaller than the corresponding single beam values.

\

\

\

\

\

\

\

\

\

\

\

{\bf This work has been submitted to the IEEE for possible 
publication. Copyright may be transferred without notice, after 
which this version may no longer be accessible.} 
 
\end{abstract}

\maketitle
\section{Introduction}
Propagation of light through atmospheric turbulence 
is the subject of intensive research owing to 
the many applications in free space laser communications \cite{Andrews98}. 
In these applications it is desirable to 
reduce the turbulence effects on the propagating light. 
One promising possibility to achieve this goal is by
using temporally partially coherent optical field consisting
of multiple laser beams with different wavelengths 
\cite{Kiasaleh2004,Kiasaleh2005,Kiasaleh2006,PM2006}. Indeed, 
in a typical setup in which the detector's response time is large 
compared with the inverse of the frequency difference between 
any pair of beams in the input field, rapidly oscillating 
contributions to the total intensity average out.
Consequently, one can expect smaller values of high moments of the 
intensity compared with corresponding single beam values. 
This would result in smaller values for the scintillation index and
for the average signal to noise ratio (SNR).

Generation of temporally partially coherent light 
consisting of multiple beams with different wavelengths 
can be efficiently realized by using 
an array of vertical external cavity surface lasers 
(VECSELs). These devices have the advantage of generating high power,   
spectrally narrow, wavelength tunable $\mbox{TEM}_{00}$  
beams (lowest order Gaussian  beams)\cite{Fallahi2006}. 

Propagation of temporally partially coherent light in atmospheric 
turbulence was first studied by Fante \cite{Fante77,Fante79}, 
who obtained approximate analytic expressions for the scintillation
index of a single infinite planar wave. 
More recently Kiasaleh studied propagation of an infinite 
multi-wavelength planar wave in weak atmospheric turbulence  
and showed that the achievable SNR is larger in the multi-wavelength case 
compared with the single-wavelength case \cite{Kiasaleh2004,Kiasaleh2005}.  
These previous studies focused on infinite planar waves, 
whereas in practice, Gaussian laser beams 
with finite initial spot size are employed. Since the dynamics of the 
optical field can strongly depend on the initial spot size 
it is important to take into account the finite spatial 
dimension of the beams. Furthermore, the optical field of $N$ 
collimated $\mbox{TEM}_{00}$ beams with different wavelengths 
depends on the wavelength separation and also on the spatial 
separation between the beam centers at the transmitter. 
Therefore, one can exploit these two different dependences to  
reduce scintillation and optimize system performance against 
turbulence effects. The dependence of the scintillation index 
on the wavelength separation of multiple overlapping 
$\mbox{TEM}_{00}$ beams was studied in Ref. \cite{Kiasaleh2006}.
It was found that for typical lasercom setups modest 
scintillation reduction of about 10$\%$ can be acheived by 
controlling the wavelength difference between the beams. We 
emphasize that only the case where the beams are completely 
overlapping at the transmitter plane was considered in 
Ref. \cite{Kiasaleh2006}. In the current Letter we focus attention 
on scintillation reduction by varying the spatial separation 
between the beams at the transmitter. We show that this 
approach leads to a much stronger decrease of the scintillation 
compared with the approach employed in Ref. \cite{Kiasaleh2006}.  
Moreover, our approach allows us to 
find the initial beam separation that minimizes scintillation, 
thus providing a simple solution 
for the important problem of optimizing temporally partially  
coherent sources of light against turbulence effects.   

In Ref. \cite{PM2006} we took the first step in this approach and 
established the framework for calculating the scintillation index
for multiple partially overlapping beams in weak atmospheric turbulence. 
Using the Rytov approximation and considering a typical 2-beam system 
we found the initial beam separation that minimizes the longitudinal 
scintillation. We showed that the longitudinal and radial scintillation 
for the optimal 2-beam configuration are smaller by about 50$\%$ 
and 35$\%$-40$\%$, respectively, compared with the corresponding 
single-beam values. However, two important aspects of the problem 
were not addressed in Ref. \cite{PM2006}: (1) the $N$-dependence of 
the longitudinal scintillation reduction compared with the single-beam case,
and (2) the possibility to optimize the system with respect to both
initial beam separation and initial spot size. In this Letter we 
address these two central issues in detail.

\section{Calculation of the scintillation index}
Consider propagation of $N$ collimated linearly polarized 
$\mbox{TEM}_{00}$ beams with different wavelengths $\lambda_{j}$, 
$j=1, ..., N$, in weak atmospheric turbulence. Assuming that the beams 
propagate along the $z$ axis and denoting by ${\mathbf d_{j}}$ 
the beam-center locations at the input plane $z=0$,  
the magnitude of the total electric field $E$  
at $z=0$ is $E({\mathbf r},0,t)=
\sum_{j=1}^{N}U_{j}({\mathbf r_{j}},0)
\exp\left[-i\omega_{j}t\right]$, 
where $U_{j}({\mathbf r_{j}},0)=
\exp[-r_{j}^{2}/W_{0j}^{2}]$, 
${\mathbf r}$ is the radius vector in the $xy$ plane, 
${\mathbf r_{j}}\equiv{\mathbf r}-{\mathbf d_{j}}$, $t$ is time, 
$k_{j}=2\pi/\lambda_{j}$ are wavenumbers, $\omega_{j}=k_{j}c$ 
are angular frequencies, and $c$ is the speed of
light. In addition, $W_{0j}$ are the initial spot sizes and we assume 
that all beams have the same amplitude.  
Assuming weak turbulence, the propagation 
is described by $N$ uncoupled linear wave equations 
\begin{eqnarray}&&
\nabla^{2}U_{j}+
k_{j}^{2}\left[1+2n_{1}({\mathbf r},z)\right]U_{j}=0,
\label{linear2}
\end{eqnarray} 
where $n_{1}({\mathbf r},z)$ represents the
refractive index fluctuations, $|n_{1}({\mathbf r},z)|\ll 1$.
To solve Eq. (\ref{linear2}) we follow Ref. \cite{Andrews98}
and employ the paraxial approximation together with 
the Huygens-Fresnel integral and the second order 
(with respect to $n_{1}$) Rytov perturbation method. 
The total intensity at $z=L$ is  
\begin{eqnarray}&&
I({\mathbf r},L,t)=
\sum_{j=1}^{N}I_{j}({\mathbf r_{j}},L)+
 \nonumber \\&&
\sum_{j}^{N}\sum_{m\ne j}^{N}
U_{j}({\mathbf r_{j}},L)
U_{m}^{*}({\mathbf r_{m}},L)
\exp\left[i(\omega_{m}-\omega_{j})t\right],
\label{int1}
\end{eqnarray}
where $I_{j}({\mathbf r_{j}},L)=|U_{j}({\mathbf r_{j}},L)|^{2}$ 
is the intensity of the $j$-th beam. 
The intensity measured by the detector is the time average
$I_{det}({\mathbf r},L)\equiv
\tau^{-1}\int_{0}^{\tau}{\rm d}t I({\mathbf r},L,t)$,
where $\tau$ is the response time of the detector. 
Assuming a slow detector and  
$\lambda_{j}\ne \lambda_{m}$ for $j\ne m$, we
neglect the terms $U_{j}U_{m}^{*}$
$j\ne m$, which are rapidly oscillating with time. 
Therefore, the measured intensity is
\begin{eqnarray}&&
I_{det}({\mathbf r},L)
\simeq \sum_{j=1}^{N}I_{j}({\mathbf r_{j}},L).
\label{int3}
\end{eqnarray}

The total scintillation index for the $N$-beam system is 
\begin{eqnarray}&&
\sigma^{2}_{I}({\mathbf r},L)=
\langle I_{det}^{2}({\mathbf r},L)\rangle
/\langle I_{det}({\mathbf r},L)\rangle^{2}-1,
\label{pcb6}
\end{eqnarray}
where $\langle\dots\rangle$ stands for
average over different realizations of turbulence disorder.
Using Eqs. (\ref{int3}) and (\ref{pcb6}) we obtain
\begin{eqnarray}&&
\sigma^{2}_{I}({\mathbf r},L)=
\left( \sum_{j=1}^{N}
\langle I_{j}({\mathbf r_{j}},L)\rangle\right)^{-2}
\left[\sum_{j=1}^{N}\langle I_{j}^{2}({\mathbf r_{j}},L)\rangle
\right.
 \nonumber \\&&
\left.
+2\sum_{j}^{N}\sum_{m>j}^{N} 
\langle I_{j}({\mathbf r_{j}},L)I_{m}({\mathbf r_{m}},L)\rangle\right]
-1.
\label{pcb6a}
\end{eqnarray}
The total scintillation index $\sigma^{2}_{I}$ can be decomposed 
into a longitudinal component
$\sigma^{2}_{I,l}(L)\equiv\sigma^{2}_{I}(0,L)$
and a radial component 
$\sigma^{2}_{r}({\mathbf r},L)\equiv
\sigma^{2}_{I}({\mathbf r},L)-\sigma^{2}_{I,l}(L)$.

In calculating intensity moments we assume that the 
perturbation field in the Rytov approximation is a Gaussian 
random variable and that the turbulence is statistically 
homogeneous and isotropic. Consequently,  
the average intensity of the $j$-th beam is given by \cite{Andrews98}
\begin{eqnarray}&&
\langle I_{j}({\mathbf r_{j}},L)\rangle=
\frac{W_{0j}^{2}}{W_{j}^{2}}
\exp\left[-\frac{2r_{j}^{2}}{W_{j}^{2}}+H_{1j}(r_{j},L)
\right],
\label{pcb7}
\end{eqnarray}
where $W_{j}$ is the spot size at distance $L$, and 
$H_{1j}$ is expressed in terms of a double 
integral of the spectral density of the 
refractive index fluctuations $\Phi_{n}(\kappa)$ over 
wavenumber $\kappa$ and propagation distance $z$. 
[See Ref. \cite{PM2006}, Eq. (18)]. The average of 
the second moment is \cite{Andrews98}
\begin{eqnarray}&&
\!\!\!\!\!\!
\langle I_{j}^{2}({\mathbf r_{j}},L)\rangle=
\langle I_{j}({\mathbf r_{j}},L)\rangle^{2}
\exp\left[H_{2j}(r_{j},L)\right],
\label{pcb9}
\end{eqnarray}
where $H_{2j}$ is given by another double 
integral of $\Phi_{n}(\kappa)$ over $\kappa$ and $z$.
[See Ref. \cite{PM2006}, Eq. (20)]. The cross-intensity term 
$\langle I_{j}({\mathbf r_{j}},L)I_{m}({\mathbf r_{m}},L)\rangle$ 
is given by \cite{PM2006} 
\begin{eqnarray}&&
\langle I_{j}({\mathbf r_{j}},L)I_{m}({\mathbf r_{m}},L)\rangle=
\langle I_{j}({\mathbf r_{j}},L)\rangle
\langle I_{m}({\mathbf r_{m}},L)\rangle\times
 \nonumber \\&&
\exp\left\{
E_{2jm}({\mathbf r_{j}},{\mathbf r_{m}};k_{j},k_{m})+
E_{2mj}({\mathbf r_{m}},{\mathbf r_{j}};k_{m},k_{j})+
\right.
 \nonumber \\&&
\left.
2{\mbox Re}\left[
E_{3jm}({\mathbf r_{j}},{\mathbf r_{m}};k_{j},k_{m})\right]
\right\},
\label{pcb13}
\end{eqnarray}
where $E_{2jm}$, $E_{2mj}$ and $E_{3jm}$ are three different 
integrals of $\Phi_{n}(\kappa)$ over $\kappa$ and $z$. 
[See Ref. \cite{PM2006}, Eqs. (22-24)].  

We consider two typical free space laser communication setups, 
in which the central 
wavelength is $\lambda_{c}=10^{-6}$m, the wavelength spacing is
$\Delta\lambda=10^{-8}$m, all beams are 
collimated and have the same initial spot size and on-axis 
amplitude. In setup A $L=1$km and the refractive index structure parameter 
is $C_{n}^{2}=3.0\times 10^{-15}\mbox{m}^{-2/3}$, and in 
setup B $L=10$km and $C_{n}^{2}=10^{-16}\mbox{m}^{-2/3}$. 
Both setups correspond to weak atmospheric turbulence conditions, 
where the Rytov variance $\sigma_{R}^{2}=1.23C_{n}^{2}k^{7/6}L^{11/6}$ 
is about 0.1 and 0.23, respectively, for all beams. 
We use the Von K\'arm\'an spectrum 
to describe the refractive index fluctuations. Thus, 
$\Phi_{n}(\kappa)=0.033C_{n}^{2}
\left(\kappa^{2}+\kappa_{out}^{2}\right)^{-11/6}
\exp\left(-\kappa^{2}/\kappa_{in}^{2}\right)$, 
where $\kappa_{in}=5.92/l_{0}$, $\kappa_{out}=1/L_{0}$,
$l_{0}$ and $L_{0}$ are the turbulence inner and outer scales, 
respectively, and $l_{0}=1.0$mm, $L_{0}$=1.0m are used.

For even $N$ we consider initial configurations in which the 
beam centers are located on a circle with diameter $d$ centered 
about the $z$-axis with equal angles between 
${\mathbf d_{j-1}}$ and ${\mathbf d_{j}}$. 
Thus, for $N=4$, for example, the centers are at 
${\mathbf d_{1}}=d{\mathbf {\hat x}}/2$, 
${\mathbf d_{1}}=d{\mathbf {\hat y}}/2$, 
${\mathbf d_{1}}=-d{\mathbf {\hat x}}/2$, and
${\mathbf d_{1}}=-d{\mathbf {\hat y}}/2$. For odd $N$ we 
consider the same geometry as in the $N-1$ case, with an 
additional beam on the $z$-axis.

\begin{figure}[htb]
\centerline{\includegraphics[width=7.5cm]{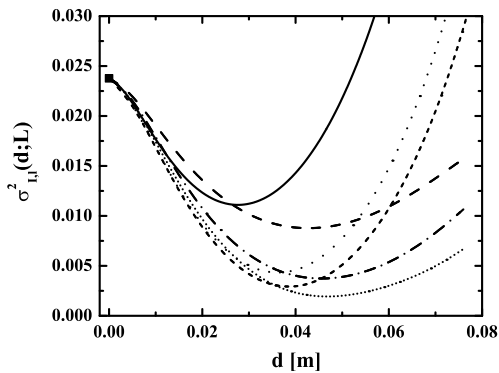}}
 \caption{Longitudinal scintillation $\sigma^{2}_{I,l}$ vs 
initial beam separation $d$ for setup A.
The solid, dashed, dotted, dashed-dotted, short-dashed, and 
short-dotted lines correspond to  2-, 3-, 4-, 5-, 8-, 
and 9-beam configurations, respectively. 
The square stands for the value for a single 
beam with the same total power and initial spot size.}
 \label{fig1}
\end{figure}

The $d$-dependence of the longitudinal scintillation for 2-, 3-, 4-, 
5-, 8-, and 9-beam configurations in setup A with initial 
spot sizes $W_{0}=1$cm is shown in Fig. \ref{fig1} 
together with the corresponding value for a single $\mbox{TEM}_{00}$ 
beam with the same total power and initial spot size. 
One can see that in each of the $N\ge 2$ cases the 
$\sigma^{2}_{I,l}$-curve exhibits a minimum at an intermediate 
$d$ value, $d_{0}=2.8$cm, 4.4cm 3.6cm, 4.6cm, 4.0cm and 4.8cm, for
the 2-, 3-, 4-, 5-, 8-, and 9-beam configurations, respectively. 
These minima correspond to the optimal configurations of the 
$N$ beams for the given physical conditions and geometric arrangements, 
where optimization is with respect to longitudinal scintillation.  
Comparison with the single-beam value shows that the longitudinal 
scintillation is reduced by 53.4$\%$, 63.2$\%$, 82.2$\%$, 
84.4$\%$, 88.1$\%$, and 92.0$\%$ for the 2-, 3-, 4-, 5-, 8-, 
and 9-beam optimal configurations, respectively. Moreover, analysis
of the longitudinal scintillation values for the optimal configurations 
shows that $\sigma^{2}_{I,l}$ decreases like $1/N$ with increasing $N$.

\begin{figure}[ht]
\centerline{\includegraphics[width=7.5cm]{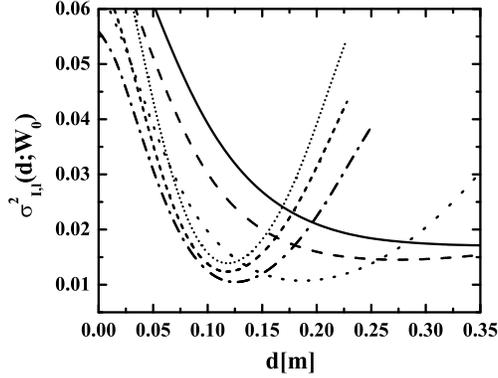}}
 \caption{Longitudinal scintillation $\sigma^{2}_{I,l}$ vs 
initial beam separation $d$ for different $W_{0}$ values 
for a 5-beam system in setup B. 
The solid, dashed, dotted, dashed-dotted, short-dashed, and 
short-dotted lines correspond to 
$W_{0}$=0.5cm, 1cm, 2cm, 4cm, 5cm, and 6cm, respectively.}
 \label{fig2}
\end{figure}

An important question concerns the possibility to further reduce 
the scintillation by optimizing with respect to 
the initial spot size. This question is addressed in 
Fig. \ref{fig2}, which shows the $d$-dependence of the longitudinal
scintillation for different $W_{0}$ values for a 5-beam system 
in setup B. One can see that the minimum value of $\sigma^{2}_{I,l}$ 
first decreases with increasing $W_{0}$ and then increases. Hence, 
the optimal configuration for the $5$-beam system in setup B, is the 
one with $W_{0}\simeq$4cm and $d_{0}=13.2$cm. Notice that the final 
free space spot size of the beams for the optimal configuration 
in this case is only 8.8cm.

\begin{figure}[htb]
\centerline{\includegraphics[width=7.5cm]{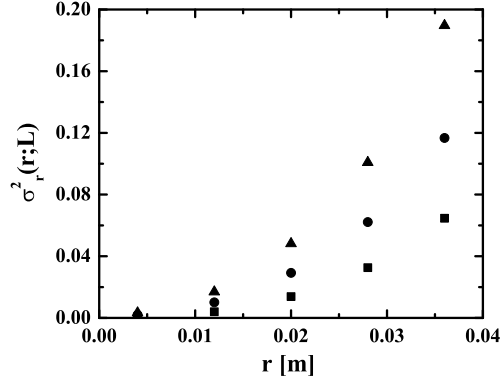}}
 \caption{Circularly averaged radial scintillation index 
$\sigma^{2}_{rr}$ vs radius $r$ for the optimal 4-beam (squares) 
and 2-beam (circles) configurations, and for a single beam with the 
same power and initial spot size (triangles), in setup A.}
 \label{fig3}
\end{figure}

When the spot size is comparable with the radius 
of the receiver's collecting lens, the radial scintillation 
becomes important. In this case it is essential 
to understand whether the radial 
scintillation of the optimal $N$-beam configurations, 
where optimization is with respect to {\it longitudinal} scintillation,
is sufficiently small compared with the single-beam value. 
Notice that in the $N$-beam case 
$\sigma^{2}_{r}({\mathbf r},L)$ is not radially symmetric. 
To enable comparison with the single-beam case we define 
the circularly averaged radial scintillation 
in the $N$-beam case as 
$\sigma^{2}_{rr}(r,L)\equiv 
\langle\sigma^{2}_{r}({\mathbf r},L)\rangle_{\theta}$,
where $\langle\dots\rangle_{\theta}$
denotes averaging over the angle $\theta$. 
Figure \ref{fig3} shows the $r$-dependence of 
$\sigma^{2}_{rr}$ for the optimal 4- and 2-beam 
configurations and for a single beam with the 
same power and initial spot size, all in setup A. 
One can see that the radial scintillation for 
the optimal 4-beam and 2-beam configurations are smaller by about
65$\%$-80$\%$ and 35$\%$-40$\%$, respectively,  
compared with the corresponding single beam values. 
Therefore, optimization of the $N$-beam configurations 
with respect to the longitudinal scintillation leads to 
significant reduction in the radial scintillation, and this reduction
effect grows with increasing $N$.

\section{Conclusion}
We calculated the scintillation index for an array of $N$ 
partially overlapping collimated 
$\mbox{TEM}_{00}$ beams with different wavelengths 
in weak atmospheric turbulence using the Rytov perturbation 
method. We showed that both 
the longitudinal and the radial scintillation can be
significantly reduced compared with the corresponding 
single-beam values by optimizing the beam array with respect to 
initial beam separation and spot size. These reduction effects 
grow with increasing $N$, resulting in a 92$\%$ reduction in the 
longitudinal scintillation for an optimal 9-beam system.

This work was supported 
by the Air Force Office for Scientific Research, Air Force Material 
Command, USAF, under grant AFOSR FA9550-04-1-0213.
JVM acknowledges financial support from the 
Alexander von Humboldt Foundation.


\begin{thebibliography}{1}
\bibitem{Andrews98} L. C. Andrews and R. L. Phillips, 
{\it Laser Beam Propagation through Random Media}  
(SPIE Press, Bellingham, Washington, 1998).

\bibitem{Kiasaleh2004} K. Kiasaleh, J. Opt. Soc. 
Am. A {\bf 21}, 1452 (2004).

\bibitem{Kiasaleh2005} K. Kiasaleh, ``Impact of turbulence on 
multi-wavelength coherent optical communications,'' in 
{\it Free-Space Laser Communications V}, D. G. Voelz and 
J. C. Ricklin, eds., Proc. SPIE {\bf 5892}, 58920R1 (2005).

\bibitem{Kiasaleh2006} K. Kiasaleh, J. Opt. Soc. 
Am. A {\bf 23}, 557 (2006).


\bibitem{PM2006} A. Peleg and J. V. Moloney, 
J. Opt. Soc. Am. A {\bf 23}, 3114 (2006).


\bibitem{Fallahi2006} L. Fan, M. Fallahi, 
J. T. Murray, R. Bedford, Y. Kaneda, A. R. Zakharian, J. Hader, 
J. V. Moloney, W. Stolz, and S. W. Koch, App. Phys. Lett. {\bf 88}, 
0211051 (2006).


\bibitem{Fante77} R. L. Fante, Radio Sci. {\bf 12}, 223 (1977). 

\bibitem{Fante79} R. L. Fante, J. Opt. Soc. Am. {\bf 69}, 71 (1979). 

\end{thebibliography}
\end{document}